\input harvmac
\input epsf
\newcount\figno
\figno=0
\def\fig#1#2#3{
\par\begingroup\parindent=0pt\leftskip=1cm\rightskip=1cm\parindent=0pt
\global\advance\figno by 1 \midinsert \epsfxsize=#3
\centerline{\epsfbox{#2}} \vskip 12pt {\bf Fig. \the\figno:}
#1\par
\endinsert\endgroup\par
}
\def\figlabel#1{\xdef#1{\the\figno}}
\def\encadremath#1{\vbox{\hrule\hbox{\vrule\kern8pt\vbox{\kern8pt
\hbox{$\displaystyle #1$}\kern8pt} \kern8pt\vrule}\hrule}}

\overfullrule=0pt

\def\half{{\textstyle{1\over2}}}

\def\half{{1\over 2}}
 
 \def\m{{\mu}}
 \def\n{{\nu}}

 \def\a{{\alpha}}
 
 \def\frac#1#2{{#1\over #2}}

 \def\D{{\Delta}}

 \def\ra{\rightarrow}
\lref\om{C. Montonen and D. Olive, Phys. Lett. B72 (1977) 117; P.
Goddard, J. Nyuts and D. Olive , Nucl. Phys. B125 (1977) 1.}

\lref\WittenZE{
  E.~Witten,
  ``Topological Quantum Field Theory,''
  Commun.\ Math.\ Phys.\  {\bf 117}, 353 (1988).
}

\lref\YamronQC{
  J.~P.~Yamron,
  ``Topological Actions From Twisted Supersymmetric Theories,''
  Phys.\ Lett.\ B {\bf 213}, 325 (1988).
}

\lref\VafaTF{
  C.~Vafa and E.~Witten,
  ``A Strong coupling test of S duality,''
  Nucl.\ Phys.\ B {\bf 431}, 3 (1994)
  [arXiv:hep-th/9408074].
}

\lref\VafaBM{
  C.~Vafa,
  ``Instantons on D-branes,''
  Nucl.\ Phys.\ B {\bf 463}, 435 (1996)
  [arXiv:hep-th/9512078].
}

\lref\VafaZH{
  C.~Vafa,
  ``Gas of D-Branes and Hagedorn Density of BPS States,''
  Nucl.\ Phys.\ B {\bf 463}, 415 (1996)
  [arXiv:hep-th/9511088].
}

\lref\MinahanVR{
  J.~A.~Minahan, D.~Nemeschansky, C.~Vafa and N.~P.~Warner,
  ``E-strings and N = 4 topological Yang-Mills theories,''
  Nucl.\ Phys.\ B {\bf 527}, 581 (1998)
  [arXiv:hep-th/9802168].
}

\lref\LabastidaIJ{
  J.~M.~F.~Labastida and C.~Lozano,
  ``The Vafa-Witten theory for gauge group SU(N),''
  Adv.\ Theor.\ Math.\ Phys.\  {\bf 3}, 1201 (1999)
  [arXiv:hep-th/9903172].
}

\lref\LozanoJI{
  C.~Lozano,
  ``Duality in topological quantum field theories,''
  arXiv:hep-th/9907123.
}

\lref\WittenEV{
  E.~Witten,
  ``Supersymmetric Yang-Mills theory on a four manifold,''
  J.\ Math.\ Phys.\  {\bf 35}, 5101 (1994)
  [arXiv:hep-th/9403195].
}

\lref\MarcusMQ{
  N.~Marcus,
  ``The Other topological twisting of N=4 Yang-Mills,''
  Nucl.\ Phys.\ B {\bf 452}, 331 (1995)
  [arXiv:hep-th/9506002].
}

\lref\WittenWY{
  E.~Witten,
  ``AdS/CFT correspondence and topological field theory,''
  JHEP {\bf 9812}, 012 (1998)
  [arXiv:hep-th/9812012].
}

\lref\GrossHE{
  D.~J.~Gross and E.~Witten,
 ``Possible Third Order Phase Transition In The Large N Lattice Gauge
  Theory,''
  Phys.\ Rev.\ D {\bf 21}, 446 (1980).
}

\lref\DabholkarJT{
  A.~Dabholkar and J.~A.~Harvey,
  ``Nonrenormalization Of The Superstring Tension,''
  Phys.\ Rev.\ Lett.\  {\bf 63}, 478 (1989).
}

\lref\DabholkarYF{
  A.~Dabholkar, G.~W.~Gibbons, J.~A.~Harvey and F.~Ruiz Ruiz,
  ``Superstrings And Solitons,''
  Nucl.\ Phys.\ B {\bf 340}, 33 (1990).
}

\lref\DabholkarYR{
  A.~Dabholkar,
  ``Exact counting of black hole microstates,''
  Phys.\ Rev.\ Lett.\  {\bf 94}, 241301 (2005)
  [arXiv:hep-th/0409148].
}

\lref\DijkgraafAC{
  R.~Dijkgraaf, E.~P.~Verlinde and M.~Vonk,
  ``On the partition sum of the NS five-brane,''
  arXiv:hep-th/0205281.
}

\lref\DijkgraafFQ{
  R.~Dijkgraaf, J.~M.~Maldacena, G.~W.~Moore and E.~P.~Verlinde,
  ``A black hole farey tail,''
  arXiv:hep-th/0005003.
}

\lref\OoguriZV{
  H.~Ooguri, A.~Strominger and C.~Vafa,
  ``Black hole attractors and the topological string,''
  Phys.\ Rev.\ D {\bf 70}, 106007 (2004)
  [arXiv:hep-th/0405146].
}

\lref\GirardelloQT{
  L.~Girardello, A.~Giveon, M.~Porrati and A.~Zaffaroni,
  ``NonAbelian strong - weak coupling duality in (string derived) N=4
  supersymmetric Yang-Mills theories,''
  Phys.\ Lett.\ B {\bf 334}, 331 (1994)
  [arXiv:hep-th/9406128].
}

\lref\GirardelloGF{
  L.~Girardello, A.~Giveon, M.~Porrati and A.~Zaffaroni,
  ``S duality in N=4 Yang-Mills theories with general gauge groups,''
  Nucl.\ Phys.\ B {\bf 448}, 127 (1995)
  [arXiv:hep-th/9502057].
}

\lref\DijkgraafBP{
  R.~Dijkgraaf, R.~Gopakumar, H.~Ooguri and C.~Vafa,
  ``Baby universes in string theory,''
  arXiv:hep-th/0504221.
}

\lref\PolchinskiUF{
  J.~Polchinski and M.~J.~Strassler,
  ``The string dual of a confining four-dimensional gauge theory,''
  arXiv:hep-th/0003136.
}

\lref\DijkgraafDH{
  R.~Dijkgraaf and C.~Vafa,
  ``A perturbative window into non-perturbative physics,''
  arXiv:hep-th/0208048.
}

\lref\GopakumarKI{
  R.~Gopakumar and C.~Vafa,
  ``On the gauge theory/geometry correspondence,''
  Adv.\ Theor.\ Math.\ Phys.\  {\bf 3}, 1415 (1999)
  [arXiv:hep-th/9811131].
}

\lref\MaldacenaBW{
  J.~M.~Maldacena and A.~Strominger,
  ``AdS(3) black holes and a stringy exclusion principle,''
  JHEP {\bf 9812}, 005 (1998)
  [arXiv:hep-th/9804085].
}

\lref\DijkgraafTE{
  R.~Dijkgraaf, S.~Gukov, A.~Neitzke and C.~Vafa,
  ``Topological M-theory as unification of form theories of gravity,''
  arXiv:hep-th/0411073.
}

\lref\HullVG{
  C.~M.~Hull,
  ``Timelike T-duality, de Sitter space, large N gauge theories and
  topological field theory,''
  JHEP {\bf 9807}, 021 (1998)
  [arXiv:hep-th/9806146].
}

\lref\deMedeirosKX{
  P.~de Medeiros, C.~M.~Hull, B.~J.~Spence and J.~M.~Figueroa-O'Farrill,
  ``Conformal topological Yang-Mills theory and de Sitter holography,''
  JHEP {\bf 0208}, 055 (2002)
  [arXiv:hep-th/0111190].
}

\lref\MaldacenaUZ{
  J.~M.~Maldacena, J.~Michelson and A.~Strominger,
  ``Anti-de Sitter fragmentation,''
  JHEP {\bf 9902}, 011 (1999)
  [arXiv:hep-th/9812073].
}

\lref\NeitzkePF{
  A.~Neitzke and C.~Vafa,
  ``N = 2 strings and the twistorial Calabi-Yau,''
  arXiv:hep-th/0402128.
}

\lref\NatsuumeQT{
  M.~Natsuume,
  ``The heterotic enhancon,''
  Phys.\ Rev.\ D {\bf 65}, 086002 (2002)
  [arXiv:hep-th/0111044].
}

\lref\BehrndtTR{
  K.~Behrndt,
  ``About a class of exact string backgrounds,''
  Nucl.\ Phys.\ B {\bf 455}, 188 (1995)
  [arXiv:hep-th/9506106].
}

\lref\KalloshYZ{
  R.~Kallosh and A.~D.~Linde,
  ``Exact supersymmetric massive and massless white holes,''
  Phys.\ Rev.\ D {\bf 52}, 7137 (1995)
  [arXiv:hep-th/9507022].
}

\lref\CveticMX{
  M.~Cvetic and D.~Youm,
  ``Singular BPS saturated states and enhanced symmetries of four-dimensional
  N=4 supersymmetric string vacua,''
  Phys.\ Lett.\ B {\bf 359}, 87 (1995)
  [arXiv:hep-th/9507160].
}

\lref\JohnsonQT{
  C.~V.~Johnson, A.~W.~Peet and J.~Polchinski,
  ``Gauge theory and the excision of repulson singularities,''
  Phys.\ Rev.\ D {\bf 61}, 086001 (2000)
  [arXiv:hep-th/9911161].
}

\lref\DabholkarDQ{
  A.~Dabholkar, R.~Kallosh and A.~Maloney,
  ``A stringy cloak for a classical singularity,''
  JHEP {\bf 0412}, 059 (2004)
  [arXiv:hep-th/0410076].
}

\lref\SenIN{
  A.~Sen,
  ``Extremal black holes and elementary string states,''
  Mod.\ Phys.\ Lett.\ A {\bf 10}, 2081 (1995)
  [arXiv:hep-th/9504147].
}

\lref\DenefRU{
  F.~Denef,
  ``Quantum quivers and Hall/hole halos,''
  JHEP {\bf 0210}, 023 (2002)
  [arXiv:hep-th/0206072].
}

\Title{hep-th/0510216}
 {\vbox{\centerline{ $S$-duality and a large $N$ phase transition}
\bigskip
\centerline{in ${\cal N} = 4$ on $K3$ at strong coupling }}}
\smallskip
\centerline{Kyriakos Papadodimas\foot{email:
papadod@fas.harvard.edu}}
\smallskip
\centerline{\it Jefferson Physical Laboratory, Harvard University}
\centerline{\it Cambridge, MA 02138, USA}\bigskip

\medskip

\noindent
We study the supersymmetric partition function of ${\cal N}= 4$
super Yang-Mills with gauge group $SU(N)$ on $K3$ in the large
$N$, fixed $g$ limit and show that it undergoes a first order
phase transition at the $S$-duality invariant value of the gauge
coupling $g$. Turning on the $\theta$-angle we find lines of phase
transitions on the $\tau$ plane. The resulting phase diagram and
the large $N$ free energy are exactly $SL(2,Z)$ invariant. Similar
phase transitions take place in systems related to the ${\cal
N}=4$ on $K3$ by dualities. One of them is the Dabholkar-Harvey
heterotic string system. We consider its mixed (a la
Ooguri-Strominger-Vafa) partition function allowing contributions
from multi-string states. We find that in the large winding charge
limit, it undergoes a phase transition with respect to chemical
potential for momentum. It is a short-string, long-string
transition that we find interesting in connection with black hole
entropy counting.

\Date{October, 2005}

\newsec{Introduction}
In this paper we study the large $N$, fixed $g$ limit of ${\cal
N}=4$   $SU(N)$ gauge theory on $K3$. Its supersymmetric partition
function was computed in $\VafaTF$ using the fact that this theory
can be twisted to a topological field theory which is simpler and
in which some quantities are exactly computable. In particular the
partition function of (physical) ${\cal N}=4$ on $K3$ is known
exactly for all values of $N$ and $g$
$\VafaTF,\MinahanVR,\LabastidaIJ$.

We show that, even though the partition function is analytic in
the gauge coupling $\tau={\theta\over 2\pi}+i{4\pi \over g^2}$ for
any finite $N$, it develops singularities as $N$ goes to infinity.
These correspond to large $N$ phase transitions with respect to
$\tau$ and, as it turns out, they are of first order. They are
sharp transitions between a phase with no instantons at weak
coupling, to a phase with a large number of instantons at strong
coupling. The resulting phase diagram is $SL(2,Z)$ invariant on
the $\tau$ plane, in agreement with $S$-duality. We find it rather
interesting that even this partition function which is protected
by supersymmetry and is independent of the metric of $K3$ (as it
corresponds to a topological observable in the twisted theory) can
undergo phase transitions.

Moreover we observe that, at least for this supersymmetric
partition function, $S$-duality becomes more manifest in the large
$N$, fixed $\tau$ limit. At finite $N$, the $S$-duality conjecture
states $\om$ that ${\cal N}=4$ with gauge group $SU(N)$ and
coupling $\tau$ is equivalent to ${\cal N}=4$ with group
$SU(N)/{\bf Z}_N$ and coupling $-{1\over \tau}$. Because the two
groups are not the same, the partition function of ${\cal N}=4$ is
not modular invariant under the full $SL(2,Z)$. However we find
that the leading piece of the large $N$ partition function is
exactly invariant under $SL(2,Z)$ indicating that $S$-duality may
act in a simpler way in this particular large $N$ limit.

We also consider the ${\cal N}=1^*$ theory on $K3$ which is
defined by a mass deformation of the ${\cal N}=4$. The phase
transition continues to exist after this mass deformation and so
the ${\cal N}=1^*$ on $K3$ undergoes a first order phase
transition with respect to the microscopic coupling $\tau$, in the
large $N$ fixed $\tau$ limit. The ${\cal N}=1^*$ theory has a
large number of isolated vacua in flat space. We argue that once
formulated on $K3$, most of these vacua become metastable and only
one of them is globally stable. For different values of the gauge
coupling, different vacua become stable. So we have phase
transitions between them as we vary the coupling.

The system that we study is related to various other systems that
can be reached by dualities starting with Euclidean $M5$ branes
wrapped on $K3\times T^2$ . The partition functions of these
systems, that count BPS states, have similar structure and we
expect similar phase transitions in the large $N$ limit, where in
this case $N$ corresponds to some charge of the system. The exact
expressions for these partition functions are known and can be
found in the literature. We will not repeat the analysis about the
existence of a phase transition for all these systems, since the
argument is almost identical with the one for the case of the
${\cal N}=4$ gauge theory.

We present the equivalent phase transition in the case of the
heterotic string, as we think it might be somehow related to the
problem of black hole entropy counting and the recent OSV proposal
$\OoguriZV$. We study the mixed partition function of the
Dabholkar-Harvey states, with fixed magnetic (winding) charge and
a chemical potential for electric charge (momentum on the $S^1$).
We include contributions from states with more than one string. We
take the large winding charge limit and find a first order phase
transition with respect to the chemical potential, between a phase
with many singly-wound strings and a phase with a long multi-wound
string. Our calculation is not exact, but we think it captures the
qualitative behavior of this system and can be made more precise
if necessary. We discuss the interpretation of this phase
transition in supergravity.

We briefly consider the dual picture of this transition in type
IIA, corresponding to the partition function of BPS states of $D4$
and $D0$-branes wrapped on $K3$.

The plan of this paper is as follows: in section 2 we discuss the
relation between the physical ${\cal N}=4$ gauge theory and its
topological version. In section 3 we present the partition
function of ${\cal N}=4$ on $K3$. In section 4 we study the large
$N$ limit of this partition function, we identify its
singularities and we discuss the interpretation of the phase
transitions in terms of instantons in the gauge theory. In section
5 we study the same transition in the ${\cal N}=1^*$ theory on
$K3$. In section 6 we turn on the $\theta$ angle and describe the
$SL(2,Z)$ invariant phase diagram of the ${\cal N}=4$ on $K3$. In
section 7 we relate this with similar transitions in systems
connected by string dualities. In section 8 we discuss the
heterotic Dabholkar-Harvey string and in 9 the $D4-D0$ system. In
sections 10 and 11 we conclude with some questions that we find
interesting.

\newsec{Topological and Physical ${\cal N} =4$}

\subsec{{\bf The topological theory}}

In this paper we will be mostly interested in the physical ${\cal
N}=4$. However the topological version of the same theory is
useful for the computation of some quantities that are protected
by supersymmetry. For more details on the topological ${\cal N}=4$
see $\LozanoJI$ and references therein.

It is well known that ${\cal N}=4$ , $SU(N)$ SYM on a 4-manifold
$M$ can be twisted to a topological theory
$\YamronQC,\VafaTF,\MarcusMQ$. This is achieved by turning on
background gauge fields for the $SU(4)$ $R$-symmetry, equal to the
spin connection of the manifold \foot{For the ${\cal N}=4$ theory
there are 3 different twistings that lead to topological theories,
depending on how we embed the rotation group $SO(4)\sim
SU(2)_L\times SU(2)_R$ into the $SU(4)$ $R$-symmetry group. Here
we are interested in the twisting considered in $\VafaTF$.}. As a
result of this, two of the supercharges become scalars and remain
unbroken even when the theory is defined on a curved manifold. The
observables that are in the cohomology of these supercharges are
topological. In particular the partition function for these models
is a topological invariant\foot{One way to compute this partition
function for K\"{a}hler manifolds is to introduce a mass
perturbation for the chiral multiplets of ${\cal N}=4$ that does
not spoil the topological nature of the theory, but which breaks
supersymmetry down to ${\cal N}=1$. This ${\cal N}=1$ theory has a
mass gap (more about it in section 5). Using the topological
invariance of the theory we can send the size of the manifold to
infinity. Then because the theory is gapped, the partition
function gets contributions only from the ground states of this
${\cal N}=1$ theory $\VafaTF,\MinahanVR$. An analogous method was
used earlier in $\WittenEV$ for the computation of the Donaldson
invariants in the topological ${\cal N}=2$ theory $\WittenZE$.}.

For the topological ${\cal N}=4$ theories (and unlike the
topological ${\cal N}=2$ $\WittenZE$) the partition function {\it
does} depend on the gauge coupling $\tau={\theta \over 2\pi} + i
{4\pi \over g^2}$. This property was used in $\VafaTF$ to test
$S$-duality for ${\cal N}=4$ at strong coupling. In the same paper
it was shown that (at least when the theory is defined on certain
manifolds, including the case of $K3$) the partition function is
the generating function for the Euler characteristic of instanton
moduli spaces for $SU(N)$ gauge theory on $M$. The partition
function has the general form:

\eqn\topologicalpart{Z_N(\tau)\sim q^{-{N \chi \over
24}}\sum_{k=0}^\infty c_{k,N}\, q^k,\qquad q=\exp(2\pi i \tau)}
where $c_{k,N}$ is the Euler characteristic of the moduli space of
$k$-instantons of $SU(N)$ on $M$, $\chi$ is the Euler
characteristic of the manifold $M$.
  For $K3$ we have $\chi=24$.

\subsec{{\bf Relation to Physical ${\cal N}=4$}}

On a general 4-manifold the twisted and the physical theories are
different. However for hyper-K\"{a}hler manifolds the twisted and
the physical ${\cal N}=4$ coincide $\VafaTF$. This means that the
(supersymmetric) partition function of the ${\it physical}$ ${\cal
N}=4$ is equal to the partition function of the twisted theory on
the same manifold, which as we said above is a topological and, in
some cases, exactly computable function for any $N$ and $g$.

    The two compact 4-dimensional hyper-K\"{a}hler manifolds are $T^4$
and $K3$. For $T^4$ the partition function is constant, so there
is not much we can say about it. On the other hand, the
topological partition function for $K3$ is nontrivial. Since $K3$
is hyper-K\"{a}hler, it is the same as the partition function of
the physical ${\cal N}=4$ on $K3$.

\newsec{${\cal N}=4$ super Yang-Mills on $K3$}
\subsec{{\bf The Vafa-Witten partition function}}

In $\VafaTF$ Vafa and Witten computed the partition function of
$SU(N)$ ${\cal N}=4$ on $K3$, for $N$ prime. Their result was
 extended to non prime $N$ in $\MinahanVR$. See also $\LabastidaIJ$.

For the gauge group $SU(N)$ with $N$ prime, the partition function
takes the form: \eqn\partitionprime{Z(\tau) = {1\over N^3} G(N
\tau) +{1\over N^2} \sum_{m=0}^{N-1} G\left({\tau+m\over
N}\right),} where \eqn\defG{G(\tau)=\eta(\tau)^{-24},}$\eta(\tau)$
is the Dedekind eta function and $\tau={\theta \over 2 \pi} + i {4
\pi \over g^2}$ the complexified gauge coupling.

For non-prime $N$ the partition function is somewhat more
complicated and is almost a Hecke transformation of order $N$ of
the function $G(\tau)$: \eqn\partitionhecke{Z(\tau) ={1\over N^3}
\sum_{0\leq a,b,d \in Z; ad=N; b\leq d-1}d\,G\left({a \tau +
b\over d}\right).} Clearly for $N$ prime $\partitionhecke$ reduces
to $\partitionprime$.

Note that $G(\tau)$ is the same\foot{Up to an overall factor of
16.} as the partition function of the Dabholkar-Harvey  BPS
heterotic string states $\DabholkarJT,\DabholkarYF,\DabholkarYR$.

One way to understand the appearance of the heterotic string
partition function is the following $\MinahanVR$ : we consider the
supersymmetric partition function of $N$ $M5$ branes on $K3\times
T^2$. We can dimensionally reduce on the $T^2$ and we end up with
${\cal N}=4$ on $K3$, whose partition function we are
studying\foot{Up to some extra factors explained in
$\MinahanVR$.}. The same partition function can be computed in the
following way: the $N$ $M5$ branes can be thought of as $N$ $NS5$
branes in IIA on $K3\times T^2$ $\DijkgraafAC$. Then by
IIA-Heterotic duality we end up with the heterotic string wound
$N$ times around the $T^2$ (all these in the Euclidean theories).

 \subsec{{\bf Some comments on $S$-duality}}

What is the behavior of $ Z(\tau)$ under $S$-duality
transformations?

The original $S$-duality conjecture for the ${\cal N}=4$ super
Yang-Mills states that the theory with gauge group $G$ and
coupling $\tau$ is equivalent to the theory with gauge group
$\hat{G}$ (the dual of $G$)\foot{The weight lattice of $\hat{G}$
is the dual of that of $G$.} and coupling $-{1\over \tau}$. The
dual group of $SU(N)$ is $SU(N)/{\bf Z}_N$, which is not simply
connected and admits 't Hooft magnetic fluxes (corresponding to
gauge bundles with fractional instanton number), unlike the
$SU(N)$ bundles for which the instanton number is integer.
Therefore when we compute the partition function of $SU(N)/{\bf
Z}_N$ we have to sum over these flux sectors. This means that in
general $Z_{SU(N)}(\tau)\neq Z_{SU(N)/{\bf Z}_N}(\tau)$. What we
expect from $S$-duality is that : \eqn\sduality{\eqalign{&
Z_{SU(N)}(\tau+1) \sim Z_{SU(N)}(\tau)\cr &
Z_{SU(N)}\left(-{1\over \tau}\right)\sim Z_{SU(N)/{\bf Z}_N}(\tau)
.}} So the partition function of $SU(N)$ is not exactly invariant
under $SL(2,Z)$ transformations. Moreover even $\sduality$ has to
be somewhat modified by assigning a nonzero modular weight to the
function $Z$, once the theory is formulated on a curved manifold.
All these issues have been analyzed in detail in $\VafaTF$. See
also $\GirardelloGF,\GirardelloQT$.

However, since we know the {\it exact} partition function of
${\cal N}=4$ on $K3$ for all $\tau$ and $N$ we can forget about
these subtleties related to $S$-duality and study how the function
$\partitionhecke$ transforms under $SL(2,Z)$ transformations of
the variable $\tau$ (without changing the group with the dual
group).

First, from the modular transformations of the function
$G(\tau)=\eta(\tau)^{-24}$ we find that the partition function
$\partitionprime,\partitionhecke$ is invariant under $\tau \ra
\tau+1$: \eqn\modulart{Z_{SU(N)}(\tau+1)=Z_{SU(N)}(\tau).} Second,
one can show (see $\LabastidaIJ$ for details) that:

\eqn\modulars{Z_{SU(N)}\left(-{1\over \tau}\right)= (N\tau)^{-12}
{1\over N^2} \sum_{a,b,d; p=gcd\{b,d\} } d^{12} p^{11} G\left({a
\tau + b \over d}\right)} We would like to emphasize that in
$\modulart,\modulars$ we are just looking at the way
$\partitionprime,\partitionhecke$ transform under $SL(2,Z)$
transformations of the variable $\tau$. These are not the actual
$S$-duality transformations, as we are not exchanging the group
$SU(N)$ with $SU(N)/{\bf Z}_N$. We see that $Z$ is not exactly
invariant under these naive $SL(2,Z)$ transformations of $\tau$,
which of course is not surprising as the $S$-duality conjecture
requires the exchange of the gauge group with its dual group and
thus the inclusion of magnetic flux sectors.

In this paper we are interested in the leading term of the
partition function of the $SU(N)$ gauge theory in the large $N$
limit. In the next section we analyze it and find that it is
exactly $SL(2,Z)$ invariant. Thus $S$-duality may act in a simpler
way in the this large $N$, fixed $\tau$ limit.

\newsec{Large $N$ limit and Phase Transitions}
\subsec{\bf The Phase Transition}

For finite $N$ the $SU(N)$ Vafa-Witten partition function is
analytic in $\tau$.  However it is possible that in the large $N$
limit the partition function develops singularities, and we can
have large $N$ phase transitions even when the theory is defined
on compact volume (an early example is $\GrossHE$). In our system
of ${\cal N}=4$ on $K3$ we will see that in the large $N$ limit
the partition function takes the form:
\eqn\asymptoticZ{Z_{SU(N)}(\tau)\sim exp\left(N
f_\infty(\tau)\right)+ {\it subleading}.} For any finite $N$ the
free energy $f_N(\tau)={\log Z_N(\tau)\over N}$ is an analytic
function of $\tau$. On the other hand, the large $N$ free energy
$f_\infty(\tau)=lim_{N\ra\infty}f_N(\tau)$ is continuous but not
everywhere smooth. This means that the system undergoes large $N$
phase transitions as we vary the gauge coupling $\tau$. \foot{We
can understand how this phase transition is possible from the
following observation. The partition function $\partitionprime$ or
$\partitionhecke$ is a sum of many terms. The argument of the
function $G$ of some of these terms goes either to infinity or to
the real axis in the limit under consideration. Using the modular
properties of $G$ and its asymptotic expansion (see Appendix A) it
is easy to show that some of these terms grow like $e^{a_i(\tau)
N}$ in the large $N$ limit. The function $a_i(\tau)$ is different
for each term. From this we conclude that the partition function
in the large $N$ limit will be dominated by the term with the
largest $Re(a_i(\tau))$. Moreover as we vary $\tau$ it is possible
that different terms (with different functions $\a_i(\tau)$)
become dominant. Thus, while each $a_i(\tau)$ are analytic, the
resulting $f_\infty(\tau)$ may have non-analytic behavior.}

For simplicity let us start with the case where the $\theta$-angle
is zero, so $\tau=i{4\pi \over g^2}$ is purely imaginary. In
appendix B we show that for $\theta=0$, there are only two terms
that can compete for dominating the partition function in the
large $N$ limit for various values of $g$. These are the terms
$G(N\tau)$ and $G\left({\tau\over N}\right)$. Their asymptotic
behavior for large $N$ is $exp\left( N {8 \pi^2 \over g^2}\right)$
and $exp\left(N {g^2 \over 2}\right)$ respectively. So in the
large $N$ limit only one of the two will dominate. Comparing them
we conclude that:

$\bullet$ At weak coupling $(g<\sqrt{4\pi})$:
\eqn\asymptoticfirst{ Z(\tau)\sim G(N \tau) \sim exp\left( N {8
\pi^2 \over g^2}\right) + ... ,}

$\bullet$ at strong coupling $(g>\sqrt{4\pi})$:
\eqn\asymptoticsecond{ Z(\tau)\sim G\left({\tau\over N}\right)
\sim exp\left(N {g^2 \over 2}\right)+... .}

The large $N$ free energy $f_\infty(g)$ is \foot{In this section
we are writing $f_\infty$ as a function of g instead of $\tau$
since we have assumed $\theta=0$.}:

$$f_\infty(g)=\cases{ {8 \pi^2\over g^2} &if  $g<\sqrt{4\pi}$;\cr {g^2 \over 2} &if $g>\sqrt{4\pi}$
.\cr}$$ So, the free energy is continuous, but its first
derivative is discontinuous at the value $g_c = \sqrt{4 \pi}$,
precisely the $S$-duality invariant value of the coupling. The
theory undergoes a first order phase transition at this value of
the coupling.

\fig{Large $N$ free energy of ${\cal N}=4$ on $K3$ as a function
of the gauge coupling $g$ for $\theta=0$.}{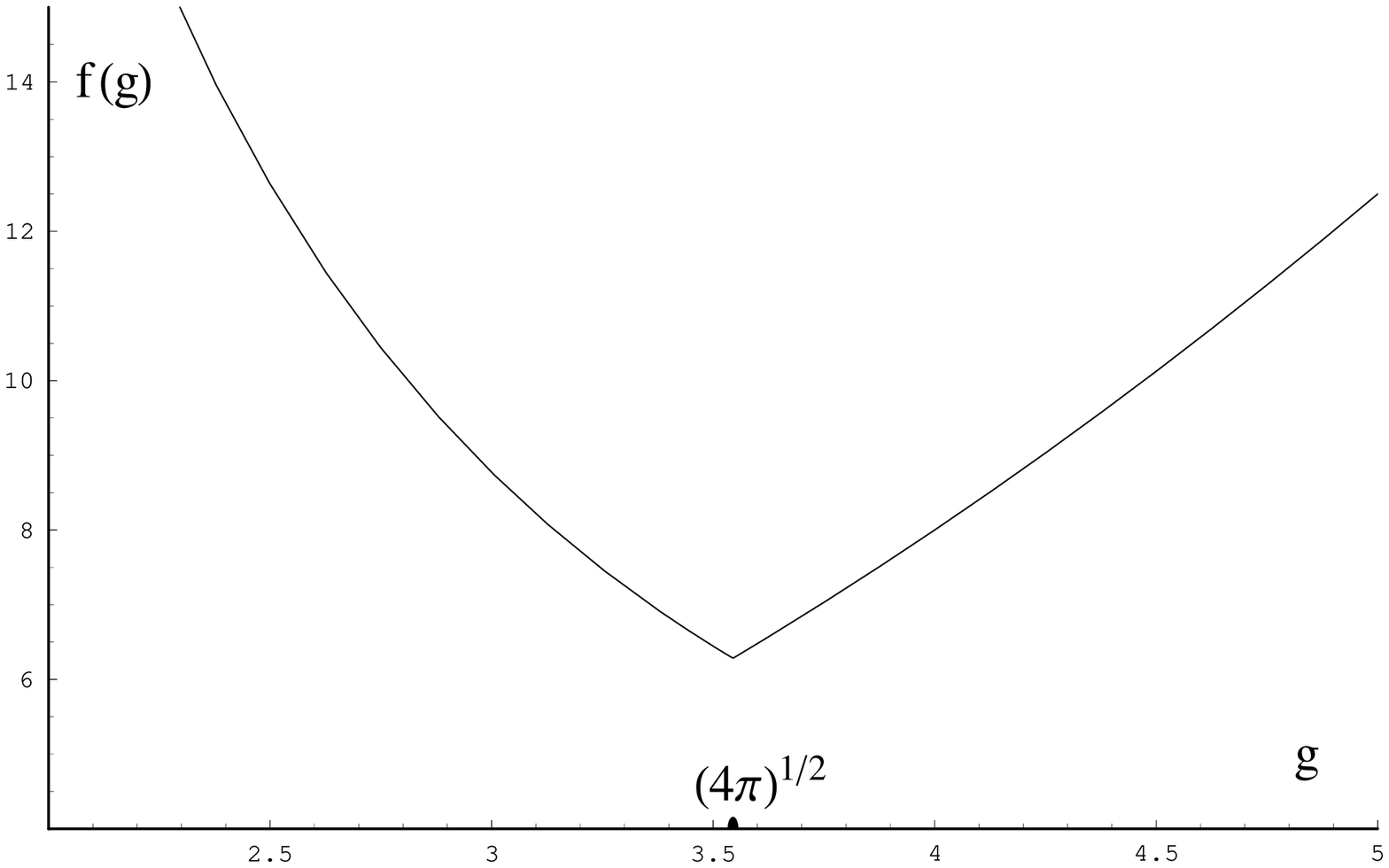}{4truein}

\subsec{\bf Physical Interpretation in terms of Instantons}

\noindent What is the interpretation of this phase transition?

\noindent According to the analysis of $\VafaTF$ the partition
function of ${\cal N}=4$ on $K3$ has the form :

\eqn\partinst{Z_N(\tau) = q^{-N} \sum_{k=0}^\infty c_{k,N} q^k}
where $q=e^{-{8 \pi ^2 \over g^2}}$ and $c_{k,N}$ is the Euler
characteristic of the $k$-instanton moduli space for $SU(N)$ on
$K3$.

We see that at weak coupling the partition function goes like
$exp\left({N 8 \pi^2\over g^2}\right)$. This means that the
partition function is dominated by the zero instanton sector at
weak coupling.

At strong coupling the partition function goes like $exp\left(N
{g^2\over 2}\right)$. At first sight this does not look like an
instanton configuration as the dependence on $g$ is of different
form from the usual $\exp\left(-k_o {8 \pi^2 \over g^2}\right)$
for some number $k_o$ of instantons. However it is possible that
the dominant configuration has instanton number that goes like
$k_o \sim N g^4$ for large $g$, explaining the $g$-dependence
\foot {I would like to thank S.Minwalla for very useful comments
and suggestions .}. If we assume that at strong coupling the
partition function is dominated by a certain $k$-instanton sector
with $k$ large, we can use an asymptotic form for $c_{k,N}$ and
try to evaluate the partition function using the saddle point
method. We assume that the Euler characteristic of the $k$
instanton moduli space for $SU(N)$ on $K3$ goes like $exp(4\pi
\sqrt{k N -N^2}) $ for large $N$ and $k$ and ${k\over N}\sim
const>1$ (see Appendix C for some details and also
$\VafaBM,\VafaZH$). We have\foot{Ignoring the contribution from
the zero instanton sector which is exponentially suppressed at
strong coupling. Also the dimension of the $k$-instanton moduli
space for $SU(N)$ on $K3$ is $\VafaTF$ : \eqn\diminstantn{dim
{\cal M}_k^N = 4 k N - 4(N^2-1)} which is negative for $1\leq k<
N$. This means that instanton sectors start at $k\geq N$ after the
trivial configuration $k=0$.} : \eqn\saddle{\eqalign{Z &\sim
\exp\left(N {8 \pi^2 \over g^2}\right)\sum_{k=N}^\infty \exp(4\pi
\sqrt{k N -N^2})\exp\left(-k{ 8 \pi^2\over g^2}\right)\cr & \sim
\exp\left(N {8 \pi^2 \over g^2}\right)\int_{x=1}^\infty \exp(N
4\pi \sqrt{x-1})\exp\left(-N x{8 \pi^2\over
g^2}\right)dx,\qquad\qquad {x={k\over N}}}} In the large $N$ limit
we find the saddle point for the integral:

\eqn\saddleb{x_{sp}={g^4\over (4\pi)^2}+1,\quad or \qquad,\quad
k_{sp} ={g^4\over (4\pi)^2} N+N} and the value of the integral is
:

\eqn\saddlec{Z_{sp}\sim \exp\left(N{g^2\over 2}\right)} which is
what we found before from the exact the partition function at
strong coupling \foot{We notice that the saddle point solution
gives $k_{sp}>N$, so we are in an instanton sector with a moduli
space of positive dimension. Also our approximation for $c_{k,N}$
is justified since $k,N$ are large and $k/N\sim const>1$.} . To
summarize we see that a possible explanation for this phase
transition is that for all values of the coupling there are two
saddle-points/phases that are competing. One is the zero instanton
phase and the other a phase with a very large number of instantons
given by $\saddleb$. At strong coupling the system is dominated by
the many-instanton phase. As we decrease the coupling the dominant
instanton number decreases and at some point the phase with no
instantons has lower free energy and we have a sharp first order
phase transition to the zero instanton sector. Figure 2 shows the
free energy diagram of the two competing phases for different
values of the coupling, and the transition at $g=\sqrt{4\pi}$.

 \fig{Free energy as a function of the dominant instanton number
$k$ for three different values of the gauge coupling
$g_1<g_2=\sqrt{4\pi}<g_3$. There are always two phases that are
competing. One is the isolated "saddle point" at $x=0$ (no
instantons), whose free energy does not depend on $g$. The other
is the minimum of the curves shown, corresponding to the saddle
point described in the text. As we increase the gauge coupling $g$
this saddle point becomes "thermodynamically" favored and at $g_2$
we have a first order phase transition between the two
phases.}{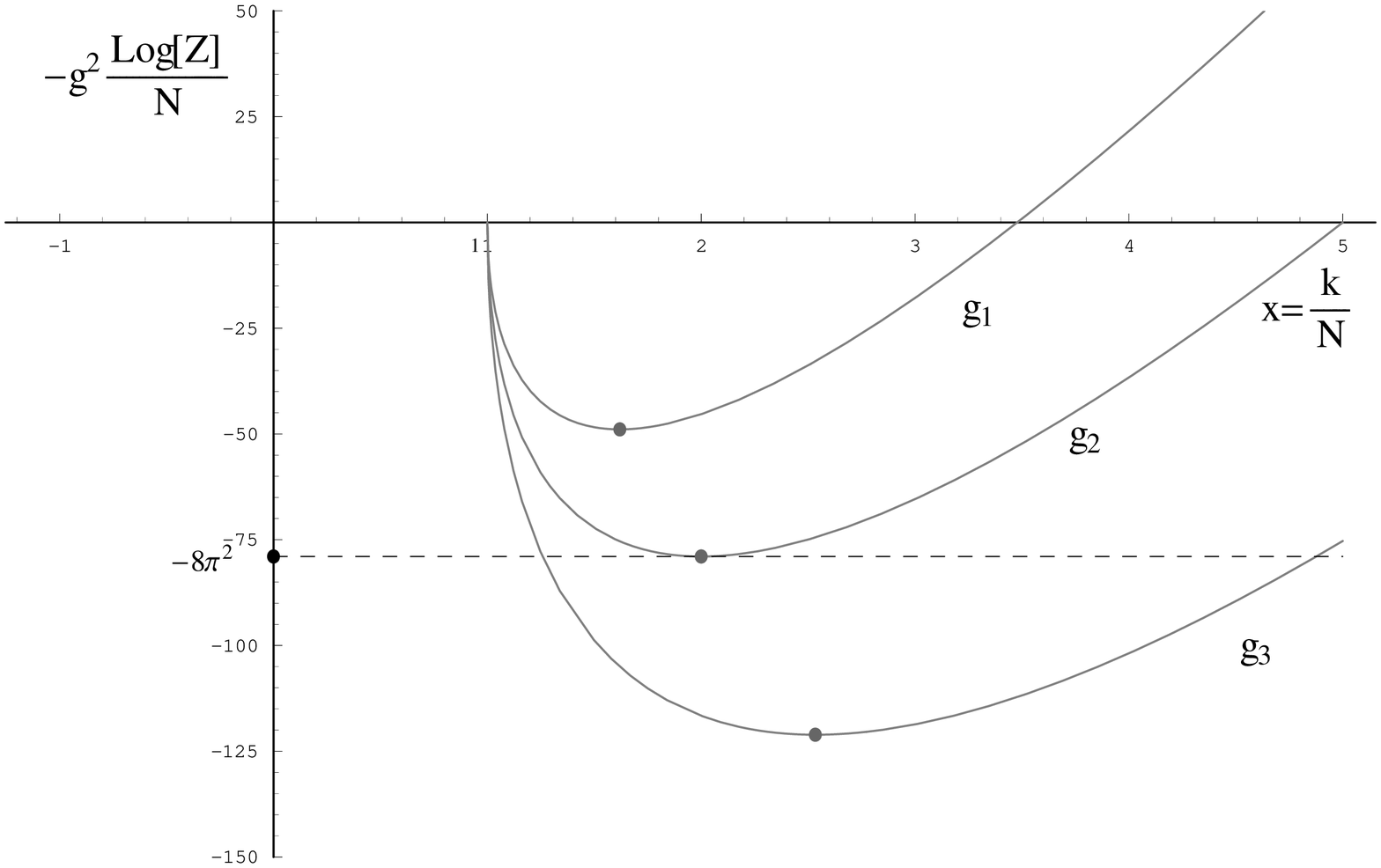}{5truein}

We can also compute the entropy of the two phases, if we assume
that the role of "temperature" is played by $g^2$:

$\bullet$ For weak coupling \foot{In this section by free energy
we mean: $F=-g^2 Log(Z)$, the thermodynamic free energy, if we
consider $g^2$ as temperature. In the other sections of the paper
we use the term free energy for $f = {Log{Z}\over N}$. We hope
that this change does not cause confusion.}(low temperature):

\eqn\weak{S_w=0,\quad E_w= - 8 \pi^2 N , \quad F_w=- 8\pi^2 N}

$\bullet$ For strong coupling (high temperature):

\eqn\strong{S_s=g^2 N ,\quad E_s= {g^4\over 2}N ,\quad F_S = -
{g^4\over 2}N}

Our conclusion is that the phase transition is between a weak
coupling phase of no instantons with small entropy and low energy,
and a strong coupling phase with many instantons, higher energy
and entropy.

\newsec{Mass deformation and the phase transition for ${\cal N}=1^*$}

It is known that the ${\cal N}=4$ theory can be deformed to the
${\cal N}=1^*$ theory by giving a mass term to its chiral
multiplets $\VafaTF,\MinahanVR,\PolchinskiUF$. We argue that the
phase transition we are studying exists also in this ${\cal
N}=1^*$ theory on $K3$ in the same large $N$, fixed $g$ limit. In
the following section we briefly describe some basic properties of
the ${\cal N}=1^*$ theory and then we discuss the characteristics
of the phase transition in ${\cal N}=1^*$.

\subsec{\bf The ${\cal N}=1^*$ theory}

The ${\cal N}=4$ super Yang-Mills in ${\cal N}=1$ language
contains a vector multiplet and 3 chiral multiplets
$\Phi_i,i=1,2,3$ in the adjoint of the gauge group, and a
superpotential:
\eqn\superpotential{W=Tr\left([\Phi_1,\Phi_2]\Phi_3\right)} We add
mass terms for the chiral multiplets to the superpotential:
\eqn\massterm{\D W= {1\over 2}m
Tr\left(\Phi_1^2+\Phi_2^2+\Phi_3^2\right)} The effect of this term
is to break supersymmetry down to ${\cal N}=1$. This theory is
called ${\cal N}=1^*$. It does not have a moduli space, but a
number of isolated vacua, which have been studied in
$\VafaTF,\MinahanVR,\PolchinskiUF$.

The perturbation $\massterm$ does not spoil the topological nature
of the theory (as it is BRST exact) on K\"{a}hler manifolds, and
there is one unbroken supercharge even after the mass
perturbation. This means that even if we turn on this mass
deformation in the ${\cal N}=4$ theory, its partition function is
independent of $m$. So the expression $\partitionhecke$ is the
(supersymmetric) partition function of ${\cal N}=1^*$ $SU(N)$ on
$K3$ for any value of $m$ and any size of the $K3$.

Let us introduce a parameter $t$ that characterizes the overall
size of the $K3$. That is, starting with an arbitrary metric
$g_{\m\n}$ on $K3$ we parametrize different metrics by $t^2
g_{\m\n}$, controlling the size of the manifold. The partition
function $\partitionhecke$ is independent of $m$ and $t$, and
exhibits a large $N$ phase transition with respect to the gauge
coupling $\tau$ for all values of $m$ and $t$.

We consider the parameter $t m$. This parameter smoothly
interpolates between the original ${\cal N}=4$ at small $t m$ and
the ${\cal N}=1^*$ {\it in its vacuum state} at large $t m$. The
phase transition with respect to $\tau$ exists for all values of
these parameters but its interpretation is different in the two
limits mentioned above.

At small $tm$ we effectively have the ${\cal N}=4$ on $K3$ and as
we saw in the previous section the phase transition is interpreted
as a transition between a configuration with no instantons at weak
coupling to a phase with many instantons  at strong coupling.

What about the limit of large $t m$?

As explained in $\WittenEV,\VafaTF,\MinahanVR$, we expect that in
this limit the partition function will be determined by the (flat
space) vacuum states of the ${\cal N}=1^*$ theory. The vacuum
structure of ${\cal N}=1^*$ has been analyzed in
$\VafaTF,\MinahanVR,\PolchinskiUF$. Classically the vacua are
given by the solutions of:
\eqn\classsvac{[\Phi_i,\Phi_j]=m\epsilon_{ijk}\Phi_k} These
equations are the commutation relations of $SU(2)$. So the
classical vacua of the theory are characterized by an
$N$-dimensional (possibly reducible) representation of $SU(2)$.

Quantum mechanically there are two categories of vacua, the
massive vacua where there are no unbroken $U(1)$ factors and there
is a mass gap, and the massless vacua. The massless vacua with the
$U(1)$ factors have extra fermion zero modes and do not contribute
$\MinahanVR$ to the supersymmetric partition function that we are
studying, so we only need to consider the vacua that have a mass
gap\foot{In the limit $N\ra\infty$ with $g$ constant, the mass gap
is of the order of the mass perturbation $m$ $\PolchinskiUF$.}.

The two extreme cases are: \item{i.}The $SU(N)$ confined vacuum:
here $\Phi_i=0$, and the unbroken group is $SU(N)$. Quantum
mechanically the theory confines, a mass gap is generated, chiral
symmetry is broken and the classical vacuum splits into $N$
quantum vacua with different expectation values for the gluino
condensate. These $N$ vacua are cyclically permuted under $\tau\ra
\tau+1$. The contribution to the partition function of ${\cal
N}=4$ on $K3$ from these vacua is given by the terms:
\eqn\contribconf{{\rm confined\,\,SU(N)\,\,vacuum}:\qquad {1\over
N^2}\sum_{m=0}^{N-1} G\left({\tau+m\over N}\right)} in
$\partitionhecke$.

\item{ii.}The other extreme case is when the $SU(N)$ group is
fully higgsed. This happens when the expectation values of the
scalars transform in the $N$-dimensional irreducible
representation of $SU(2)$. Then a mass gap is generated even
classically. The contribution to the partition function from this
vacuum is:

\eqn\contribhiggs{{\rm fully\,\,higgsed\,\,vacuum}:\qquad {1\over
N^3}G(N\tau)}

There are other intermediate {\it massive} vacua when $N$ has
divisors. If for example $N=a d$, there is a vacuum where the
expectation values of the $\Phi's$ transform in $d$ copies of the
$a$-dimensional irreducible representation of $SU(2)$. Then the
unbroken gauge group is $SU(d)$. Quantum mechanically it confines
and splits into $d$ vacua. The contribution from these vacua is:

\eqn\contribinterm{{\rm intermediate\,\,vacua}:\qquad{1\over N^3}
\sum_{b=0,..d-1}d G\left({a\tau +b\over d}\right)}

So as explained in $\MinahanVR$ we see that all massive vacua of
the ${\cal N}=1^*$ theory are in one to one correspondence with
the various terms of the Hecke transformation that give the
partition function $\partitionhecke$.

All these vacua are interchanged by the action of $S$-duality in
${\cal N}=4$.

\subsec{\bf Phase transition in ${\cal N}=1^*$}

We saw above that in flat space the ${\cal N}=1^*$ theory has a
number of (supersymmetric) massive vacua, all of course with zero
energy.

Now let us consider the same theory on compact volume, starting
with the example of a flat $T^4$. Let us choose a small region of
the $T^4$ and look at the fields in that region. We can send $t$
to infinity\foot{Again, $t$ parametrizes the overall size of the
manifold.}, blowing up the size of this region as much as we want
without changing the partition function. For very large $t$ and
because the theory has a mass gap \foot{As explained in
$\MinahanVR$ vacua without mass gap do not contribute to the
partition function because of extra fermionic zero modes.} we
expect to find the fields in this region in their (flat space)
vacuum configuration. But which of the many vacua of ${\cal
N}=1^*$? Since we are in compact volume we can't have spontaneous
symmetry breaking (at finite $N$). The wavefunction of the system
is not localized around one vacuum only, but is in superposition
of the various vacua.

Now consider the theory on $K3$. As in the case of $T^4$ the
partition function receives contributions from all (flat space)
vacua as we can see from the expression $\partitionhecke$. However
there is a difference. We notice that because of curvature
corrections the contributions from different vacua are not equal
and depend on the coupling constant $\tau$.\foot{These are the
gravitational corrections that correspond to the genus 1 diagrams
in the Dijkgraaf-Vafa matrix model around the corresponding vacuum
of ${\cal N}=1^*$ $\DijkgraafDH$.} Notice that these contributions
are not extensive in the volume of $K3$ (as they are independent
of $t$), and they only depend on its topology . So once we are on
$K3$ these curvature corrections make most of the (flat space)
vacua become metastable and precisely which one is globally stable
depends on the value of the coupling $\tau$. But since we are in
finite volume and at finite $N$ we still get contributions from
all of them \foot{As in any quantum system in compact volume with
metastable vacuum states. In our system the energy difference
between the vacua is not proportional to the volume $t$ of the
$K3$, so when we send $t$ to infinity we do not have a decay to
the stable vacuum. On the other hand the energy difference is
proportional to $N$ so in the large $N$ limit we have the
localization of the wavefunction of the system around the
absolutely stable vacuum, and phase transitions as we change the
coupling.}, and this is reflected in the form of
$\partitionhecke$.

In the large $N$ limit spontaneous symmetry breaking can take
place even in finite volume. Even if there is a very small
difference in "free energy" between two metastable vacua, if this
difference scales like $N$ then in the large $N$ limit  the theory
jumps to the vacuum with the smaller free energy, depending on the
value of the coupling. The contributions from other vacua are
exponentially suppressed. For example for $\theta=0$: at weak
coupling $g$ the most stable vacuum is the fully higgsed
(corresponding to the term $G(N\tau)$). At strong coupling the
confined $SU(N)$ vacuum (corresponding to $G\left({\tau\over
N}\right)$ becomes more stable and we have a first order phase
transition between the two at $g=\sqrt{4\pi}$.

\fig{Phase Diagram of ${\cal N}=1^*$ $SU(N)$ on $K3$. We see the
phase transition line at $g=\sqrt{4\pi}$ for all values of $tm$.
At large $tm$ the most natural physical interpretation of the
phase transition is as a first order transition between two
different "ground states" of ${\cal N}=1^*$ on $K3$. At small $tm$
the theory looks like ${\cal N}=4$ on $K3$ and as we concluded
before the phase transition is between a phase with no instantons
to a phase with many.}{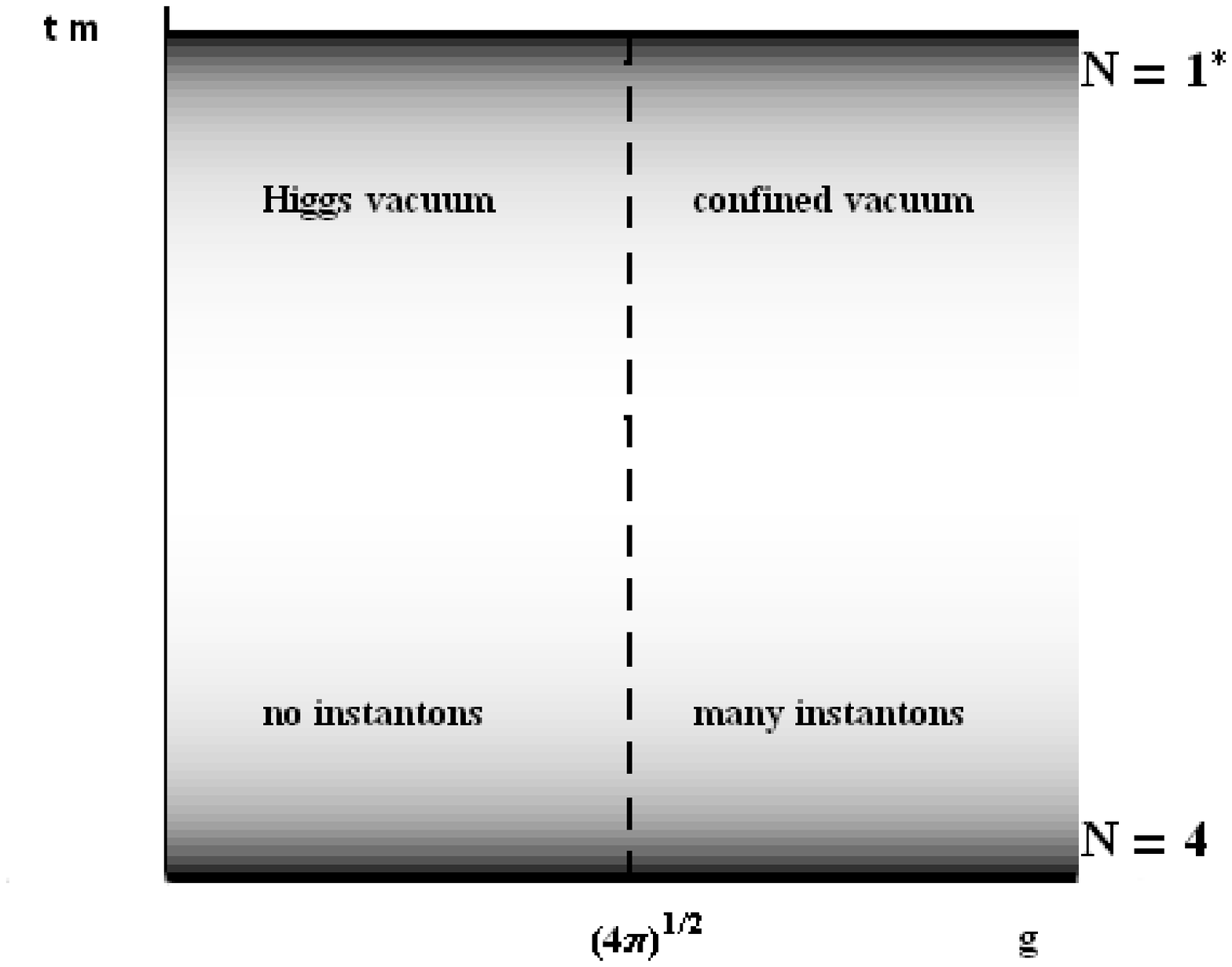}{4truein}

To summarize, for any value of the coupling there are many
metastable vacua for ${\cal N}=1^*$ on $K3$. At infinite $N$ the
theory selects the most stable vacuum, but which one it is,
depends on the value of the coupling. As we change the coupling
different competing metastable vacua become absolutely stable and
we have first order phase transitions between them. At finite $N$
the theory is in a superposition of all vacua, as we can see from
the $\partitionhecke$.

In figure 3 we see the phase diagram as a function of the gauge
coupling $g$ and the parameter $t m$ that controls the amount of
breaking of the ${\cal N}=4$ supersymmetry. In the two limits of
large and small $t m$ the interpretation of the phase transition
is more clear. For intermediate values of $t m$ the phase
transition still happens but is more difficult to interpret
physically.

\newsec{Turning on the $\theta$ angle and an $SL(2,Z)$ invariant
phase diagram}

We now turn to the case of non-zero $\theta $. Again for every
value of the coupling $\tau$ we have to identify the term that
dominates the partition function, compute the large $N$ free
energy $f_\infty(\tau)$ and study its singularities.

This can be easily done by using the modular properties
$\modulart$ and $\modulars$ of the function $Z_{SU(N)}(\tau)$. We
make the following observation: we know (see Appendix B) that for
any $\tau$ in the first fundamental domain of $SL(2,Z)$:
\eqn\firstdomain{{\cal F}_o=\{ |\tau|>1,\quad
|Re(\tau)|<\half,\quad Im(\tau)>0 \}} the large $N$ partition
function is dominated by the term $G(N\tau)$, while all others are
(exponentially) suppressed compared to this largest term. From
this observation and $\modulart,\modulars$, we conclude that for
two different values of $\tau$ which are related by an $SL(2,Z)$
transformation, the dominant term has the same value or more
precisely the same exponential growth\foot{The prefactor of the
exponential will be different in general, but it is at most
polynomial in $N$ and drops out when we compute the large $N$ free
energy.} with $N$.

This has two interesting consequences. First, it means that we
only need to study the large $N$ limit of the partition function
in the first fundamental domain ${\cal F}_o$ and then extend it
everywhere else on the plane by $SL(2,Z)$ transformations. Second,
the same property means that in the large $N$ limit the free
energy becomes exactly invariant under $S$-duality transformations
without the exchange of the gauge group with its dual group.

In the first fundamental domain it is always the term $G(N\tau)$
that dominates, giving the free energy:
\eqn\freefirst{f_\infty(\tau)=-2 \pi i \tau ,\quad \tau\in {\cal
F}_0} and from our arguments above the function must be extended
everywhere else on the plane by $SL(2,Z)$ transformations, as if
it were a modular form of weight 0. That is:
\eqn\freeplane{f_\infty({\tau})=-2 \pi i {a \tau +b \over c\tau
+d}} where $\left(\matrix{a&b\cr c&d}\right)$ is the unique
$SL(2,Z)$ transformation that maps any given $\tau$, inside the
first fundamental domain ${\cal F}_o$.

So the conclusion is that the large $N$ free energy transforms
like a modular form of weight 0 under $SL(2,Z)$ transformations.
Thus we see that $S$-duality acts in a simpler way in the theory
in the large $N$ limit.

\fig{Phase Diagram of ${\cal N}=4$ $SU(N)$ on $K3$, in the large
$N$ fixed $\tau$ limit. The lines correspond to first order phase
transitions.}{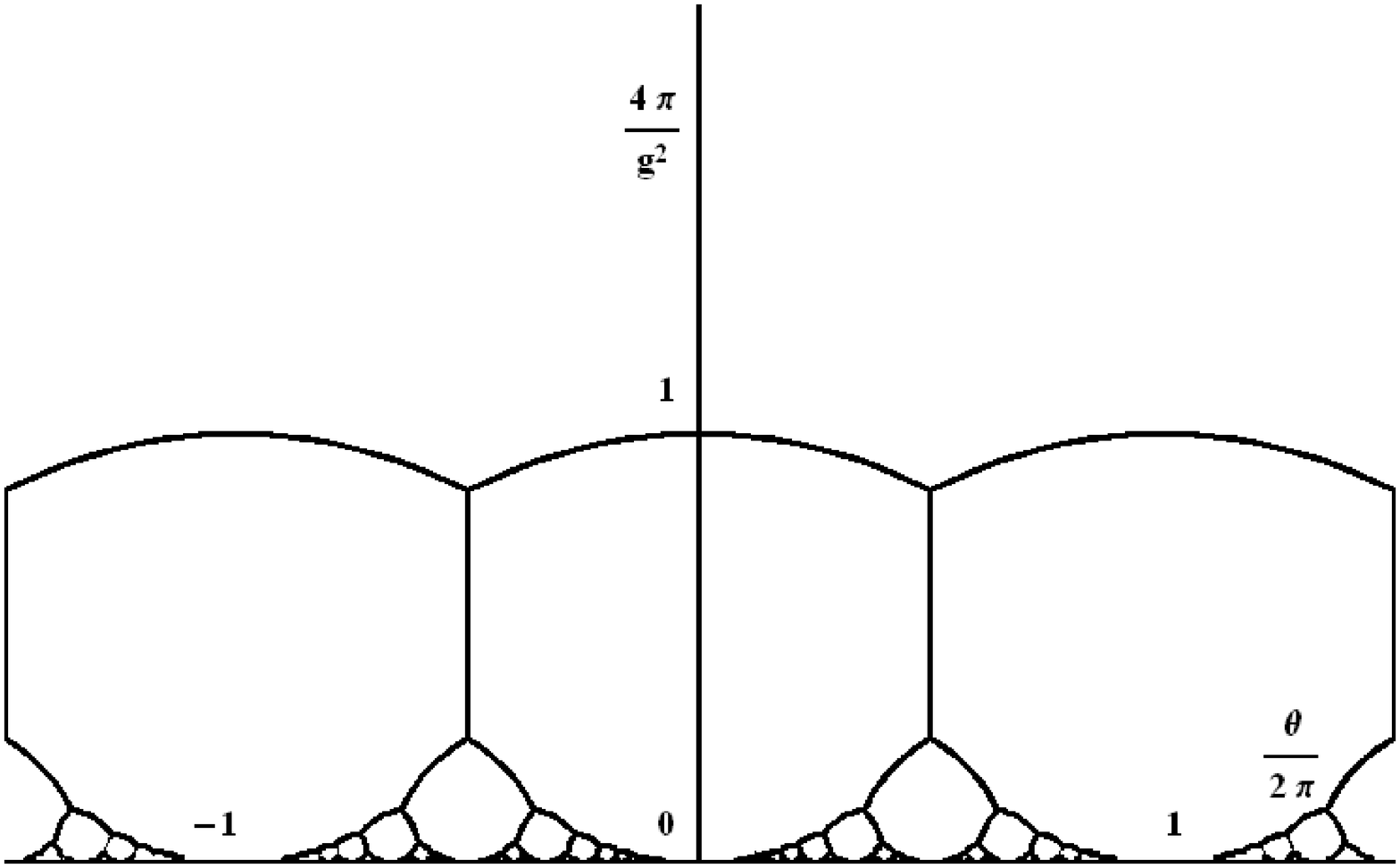}{5truein}

However, as we saw in the previous section the free energy it is
not necessarily analytic. It is easy to see that this large $N$
free energy $f_\infty(\tau)$ is continuous everywhere\foot{There
are some trivial discontinuities along the vertical boundaries of
the first fundamental domain ${\cal F}_o$ and all its images under
$SL(2,Z)$, on which the imaginary part of the free energy jumps by
$2\pi$. Of course these are not real physical singularities as the
partition function $Z\sim exp(N f_\infty)$ is continuous under
this jump of the imaginary part of $f_\infty$.}, but its first
derivative with respect to the coupling is discontinuous along the
lower boundary of the first fundamental domain $|\tau|=1,\,
|Re(\tau)|<\half$, as well as on its images under all $SL(2,Z)$
transformations. So we have first order phase transitions with
respect to the gauge coupling $\tau$ along these lines. In figure
4 we can see these lines of first order phase transitions, that
characterize the $SL(2,Z)$ invariant phase diagram for ${\cal
N}=4$ on $K3$. For a similar phase diagram see
$\MaldacenaBW,\DijkgraafFQ$.

\newsec{Connections with other systems}

There are BPS systems of D-branes, related by dualities, whose
supersymmetric partition function has a similar structure: it is
an order $N$ Hecke transformation of some smooth function of
$\tau$. $N$ corresponds to some charge of the system and $\tau$ to
a chemical potential or complex modulus of a $T^2$ in spacetime.
The partition function is a smooth function of $\tau$ for finite
$N$ but it develops singularities as $N$ goes to infinity . These
systems will exhibit similar first order phase transitions with
respect to $\tau$. We mention some of them:

$\bullet$ $N$ (Euclidean) $M5$ branes in $M$ theory wrapped on
$K3\times
 T^2$. This system was discussed in $\MinahanVR$. Its partition
function is an (almost) Hecke transformation of order $N$ of some
smooth modular form. In the large $N$ limit we expect a phase
transition with respect to the complex modulus $\tau$ of the $T^2$
, similar to the one we studied in this paper.

$\bullet$ Similarly for $N$ $NS5$ branes in type IIA wrapped on
$K3\times T^2$. This system was studied in $\DijkgraafFQ$ and the
story is similar with the cases above.

$\bullet$ The low energy world volume theory of these $N$ $M5$
branes is the $6D$ theory $A_{N-1}$ SCFT on $K3\times T^2$. These
theories (on flat space) have no parameters. From the above we
conclude that if we formulate the $A_{N-1}$ theory on $K3\times
T^2$ and  send $N$ to infinity, we expect a first order phase
transition as a function of the modular parameter $\tau$ of the
$T^2$.

$\bullet$ In type IIA the partition function for BPS states of $N$
$D4$ branes wrapped on $K3$ and with chemical potential $\tau$ for
$D0$ brane charge and by $T$-duality, in  IIB the partition
function for BPS states of $D1-D5$. See also
$\MaldacenaBW,\DijkgraafFQ$.

$\bullet$ Finally as explained in $\MinahanVR$, by IIA-Heterotic
duality we have the same structure for the partition function
which counts BPS states for the heterotic string wound $N$ times
on a $T^2$. (Note that this is in Euclidean signature, so we must
be careful about the interpretation of the multi-wound heterotic
string). The partition function depends on the complex parameter
$\tau$ of the $T^2$. In the large $N$ limit we find first order
phase transitions with respect to $\tau$. The different phases
correspond to configurations where the heterotic string wraps a
specific cycle of the $T^2$ $N$ times. As we change $\tau$ this
distinguished cycle changes, making the large $N$ partition
function non-smooth.

\newsec{Heterotic String Transition}

We find the case of the multi wound heterotic string rather
interesting in connection with black hole entropy counting and the
Ooguri-Strominger-Vafa proposal $\OoguriZV$. According to the OSV
prescription the partition function of the black hole has to be
computed with fixed magnetic charge, arbitrary electric charge and
with some chemical potential for electric charge. Motivated by the
existence of the phase transition studied above, we will consider
the partition function of the Dabholkar-Harvey states of the
heterotic string with fixed magnetic charge $W$ (winding) and
fixed chemical potential $\phi_0$ for the electric charge $P$
(momentum on the $S^1$). We will include contributions from states
that consist of more than one strings. We find a phase transition
with respect to the chemical potential $\phi_0$ in the large $W$
limit.

It is well known that the heterotic string compactified on an
$S^1$ has an infinite tower of massive BPS states
(Dabholkar-Harvey states). These are states of winding $w$ and
momentum $p$. The right moving (supersymmetric) sector of the
heterotic string is in its ground state, while the left moving
sector can be excited to an arbitrary oscillator level $N_L$. The
only constraint is the level matching condition:

\eqn\levelmatching{N_L  = 1-p w} All of these states are BPS
$\DabholkarJT,\DabholkarYF,\DabholkarYR$.

We want to compute the partition function of these states with
fixed total winding charge $W$ and chemical potential $\phi_0$ for
total momentum $P$ on the $S^1$, {\it allowing contributions from
multi-string states}\foot{Since we are interested in BPS
configurations we have to take all the charge vectors $(p_i,w_i)$
of the various strings to be parallel.} . So we want to compute:
\eqn\partitionheterotic{Z_W(\phi_0)= \sum_{BPS\,states} e^{2 \pi
\phi_0 P}} To illustrate our point we will only consider the
contributions from two kinds of states. The exact computation
including all possible configurations with total winding $W$ can
be done if necessary.

$\bullet$ Consider the configuration of a single multi-wound long
string ($w=W$). As we see from $\levelmatching$ the momentum $p$
can take the values $p=0, -1, -2, ...$ but not $p=1$ if $W$ is
large and positive, since $N_L\geq 0$. For large $W$ the partition
function can be computed using the saddle point method\foot{The
degeneracy of a single-string state with momentum p and winding w
grows like $\exp(4\pi \sqrt{|p w|})$} and we find:
\eqn\singly{Z^{long}_W\sim e^{2 \pi{1\over \phi_0}W}}the electric
charge for this configuration at the saddle point is
:\eqn\longp{P^{long}=-{W\over \phi_0^2}}

$\bullet$ Now consider the configuration of $W$ singly-wound
strings, so for each of them $w=1$. The momentum of each string
can take the values $p=1, 0, -1 ,-2....$. We will consider only
the contribution from the "ground state"\foot{In the large $W$
limit, the contribution from other states is suppressed.} $p=1$.
Since we have $W$ strings $P_{total}=W$. The degeneracy for this
state is $1$ since $\levelmatching$ gives $N_L=0$ and, as we
discuss below, we believe that because of the BPS condition there
is no contribution to the entropy from the transverse
volume\foot{Except for the center of mass volume factor that is
common in both phases and which we do not need to include when we
compare the two configurations.}. This means that the contribution
of $W$ singly wound $(w=1,p=1)$ strings to the partition function
is: \eqn\singly{Z^{short}_W\sim e^{2 \pi \phi_0 W}} and the
electric charge:\eqn\singlyp{P^{short}=W}
 So the partition function for total winding charge $W$
and chemical potential $\phi_0$ for momentum will be:
\eqn\total{Z_W(\phi_0) =Z^{short}_W+Z^{long}_W+...} or
\eqn\totalb{Z_W(\phi_0) \sim  \exp(2 \pi \phi_0 W)+\exp\left(2 \pi
{1\over \phi_0}W\right)+...} where we have not written the
contributions from other configurations. But just from $\totalb$
we see that in the large $W$ limit, the system will undergo a
first order phase transition at the value of the chemical
potential $\phi_0=1$. We believe that the exact counting and the
inclusion of the other states will not change this result
qualitatively.

One might worry that in the multi-string phase we have to include
the entropy factors from the volume transverse to the $S^1$. We
think that since we are interested in a BPS configuration there is
only one allowed state of the $W$ short heterotic strings. To
understand this better we ask how our configurations look in the
supergravity approximation. Clearly the one-long-string looks like
a small BPS heterotic black hole. Classically the horizon area is
zero but as discussed in $\SenIN,\DabholkarYR,\DabholkarDQ$ we
expect that stringy corrections will generate a horizon. Computing
the entropy from supergravity with corrections gives the same
answer with the microscopic counting. On the other hand for the
many-short-string phase we do not have a black hole solution,
since in this configuration each of the $W$ heterotic strings has
unit winding and unit momentum, so from $\levelmatching$ the
oscillator level for the left side is $N_L=0$. The usual BPS
heterotic small black holes correspond to winding and momentum
charges $(W,P)$, with $W$ and $P$ large and $P W<0$, so $N_L\geq
1$. However, there are BPS solutions of the low energy
supergravity $\BehrndtTR,\KalloshYZ,\CveticMX$ with the desired
charges $(W,W)$, so now $P W = W^2>0$, corresponding to a
collection of BPS heterotic strings with $N_L=0$. Naively these
solutions have timelike naked singularities but as explained in
$\JohnsonQT$ they are resolved by the enhan\c{c}on mechanism. The
resulting geometry is regular with no horizon. It is BPS and
carries the same charges with the collection of the many
singly-wound heterotic strings. So we find it plausible that the
many-short-string-phase is described in supergavity by the
"heterotic enhan\c{c}on" solution $\NatsuumeQT$. This is
consistent with our assumption that the number of states in this
phase is 1 giving zero entropy, since this solution does not have
a horizon\foot{We could also consider the zero string coupling
limit and place the strings in a box of volume $V$ (if we could
somehow ignore the issue of having nonzero net charge in compact
volume). Then we see that there is only one BPS state: the one
where all strings are completely delocalized in the box. Trying to
localize any of them would introduce additional 'zero-point'
energy and lift the energy of the state above the BPS bound, so
naively no volume factors have to be included in the entropy. The
phase transition point is independent of the volume $V$, which we
can then send to infinity.}. We think it is interesting to
investigate this configuration in possible connections with multi-centered
black holes, see also
$\MaldacenaUZ,\DenefRU,\OoguriZV,\DijkgraafBP$.

To summarize we find a first order "phase transition" between a
configuration of many singly-wound strings and a single long
multi-wound string. Both configurations are BPS. In analogy with
the ${\cal N}=4$ gauge theory that we studied before, it is a
transition between a phase with low energy and zero entropy (short
strings) to a phase with large energy and large entropy (long
string). We notice from $\longp$ and $\singlyp$ that the value of
the electric charge $P$ is different in the two phases. In
supergravity the two configurations correspond to a smooth
solution without a horizon(short strings) and to a small BPS black
hole (long string).

We find it interesting that the partition function we considered
has an OSV form, since it is computed at fixed magnetic charge $W$
and with chemical potential $\phi_0$ for electric charge. One
might expect that the OSV index defined on a boundary CFT will
receive contributions from both configurations and exhibit the
phase transition mentioned above.

\newsec{\bf $D4/D0$ system}

We briefly consider the dual system in type IIA, which is a number
of $D4$ branes wrapped on $K3$ with $D0$ branes on them. It is
well known that a $D4$-brane wrapped on $K3$ acquires an induced
$D0$-brane charge equal to $-1$. So, for $N$ $D4$ and $k$
$D0$-branes the total $D0$ charge is $Q=k-N$. We want to compute
the partition function of BPS states with fixed $D4$-brane charge
$N$ and chemical potential $\phi_0$ for $D0$-brane charge $Q$:
\eqn\dfdz{Z_N(\phi_0)=\sum_{BPS\,states} exp\left(-2\pi \phi_0
Q\right)}

$\bullet$ One kind of configurations is that of $N$ $D4$ and $k>0$
$D0$ branes forming a bound state that looks like a small BPS
black hole in type IIA. The contribution from these configurations
is: \eqn\bounddf{Z_{BH}(\phi_0) = \sum_{k} d_{N,k} exp\left(-2\pi
\phi_0 (k-N)\right)} where $d_{N,k}$ is the number of bound states
of $N$ $D4$-branes and $k$ $D0$-branes. But we know $\VafaBM$ that
for large $N$ and $k$ we have $d_{N,k} \sim exp\left(4\pi \sqrt{N
k -N^2}\right)$. Following the analysis of the previous sections,
we compute the partition function in the large $N$ limit by a
saddle point approximation: \eqn\boundsaddle{Z_{BH}(\phi_0) \sim
\exp\left({2 \pi \over \phi_0} N\right)} Also on the saddle point
we find that the $D0$ brane charge for the bound system is equal
to : \eqn\bounddzcharge{Q_{BH}= {N\over \phi_0^2}} Some of it is
due to the induced $D0$-brane charge on the $D4$-branes, so the
number of $D0$-branes is: \eqn\numerofdz{k={N\over \phi_0^2}+N}

$\bullet$ The second configuration is that of $N$ $D4$-branes,
with no $D0$-brane. In this case the corresponding supergravity
solution is the enhan\c{c}on geometry $\JohnsonQT$, which is
smooth and has no horizon. Thus the entropy is zero, or there is
only one state. The only $D0$-brane charge is the induced one, so:
\eqn\unbounddzchare{Q_{enh}=-N}And we have:
\eqn\enh{Z_{enh}(\phi_0)\sim \exp\left(2\pi \phi_0 N\right)}

Comparing $\boundsaddle$ and $\enh$ we see that  in the large $N$
limit there is a phase transition at $\phi_0=1$ between the two
configurations. In the supergravity description it is a first
order transition between an extremal black hole and the
enhan\c{c}on solution for the $D4/D0$ system. This result in IIA is
in agreement with the $S$-dual system of the heterotic string that
we studied in the previous section. Again we notice that the two
saddle points have the same value of $D4$-brane charge but
different values of $D0$-brane charge.

\newsec{Holographic Duals}

The phase transition in ${\cal N}=4$ on $K3$ that we studied may
be related to string theory in yet another way.  As for ${\cal
N}=4$ on flat space or on $S^3\times R$, in the large $N$ limit we
expect some kind of dual string theory, like in the usual AdS/CFT
correspondence. If such a dual string theory exists for ${\cal
N}=4$ on $K3$ then we can tell that it will have a phase
transition in the large $N$ fixed $g_{YM}$ limit, as a function of
$g_{YM}$. Of course we do not know the exact correspondence
between the parameters of ${\cal N}=4$ on $K3$ and its holographic
dual string theory, but if we assume that it is similar to the
standard AdS/CFT we conclude that this phase transition on the
string side will take place in the limit of locally flat space
($N\ra \infty$) and finite string coupling ($g_s \sim g_{YM}^2\sim
const$). The phase transition is with respect to the string
coupling.

We also observed that the large $N$ free energy was exactly
invariant under $S$-duality. Let us consider for a moment the
${\cal N}=4$ theory on any other manifold, and let's look at its
partition function, possibly with supersymmetry breaking boundary
conditions (thermal partition function). It is reasonable to
assume that in the large $N$, fixed $g$ limit the partition
function will have the form: \eqn\generalp{Z_N(g) \sim \exp(N^a
f(g)) + ...} for some $a$. Now if we assume that  $f$ is exactly
invariant under $S$-duality, as it happened in our system, we
have: \eqn\dualitys{f\left({4\pi\over g}\right)= f(g)} We conclude
that the first derivative is either discontinuous at the self-dual
point $g=\sqrt{4\pi}$ or exactly zero $f'(\sqrt{4\pi})=0$. In the
system we studied we found that the first possibility
(discontinuity) is realized. Even though we have no indication for
it, it would be rather fascinating if the same thing happened for
example in ${\cal N}=4$ on $S^3\times S^1$ in the large $N$ fixed
$g$ limit\foot{If the theory is defined on $M=S^4$ or $M=S^3\times
S^1$, the $SU(N)$ and $SU(N)/{\bf Z}_N$ gauge groups give
essentially the same results, as $H^2(M,{\bf Z}_N)$=0 for these
manifolds and there are no magnetic flux sectors. This indicates
that $\dualitys$ must be true in these cases.}. Such a
possibility, if true, would imply a phase transition for $IIB$
string theory in the flat space, finite string coupling limit.

\newsec{Conclusions}
We have demonstrated that ${\cal N}=4$ and ${\cal N}=1^*$ on $K3$
exhibit first order phase transitions with respect to the gauge
coupling $g$, in the large $N$ fixed $g$ limit.

There are several aspects of these phase transitions that we find
interesting:

\item{i.}They take place in the large $N$, fixed $g$ limit instead
of the usual 't Hooft limit. In the standard AdS/CFT this
corresponds to the limit of (locally) flat space and finite string
coupling. One then expects that if we knew a holographic dual
string theory of ${\cal N}=4$ on $K3$ we would have a phase
transition in the string theory at finite string coupling.

\item{ii.}They are phase transitions affecting the supersymmetric
partition function, which is usually considered a protected
quantity that does not change (or varies smoothly) with respect to
continuous changes of the parameters of the theory.

\item{iii.}The same phase transition may exist for the topological
or the physical ${\cal N}=4$ on other manifolds. Of course it
would be really exciting if such a phase transition existed in the
physical ${\cal N}=4$ at finite temperature on $S^3$ or $R^3$ as
this would imply the same for type IIB string theory at finite
string coupling on flat space.

\item{iv.}It would be interesting to study more carefully the
interpretation and implications of the analogous phase transitions
in the systems of five-branes, heterotic strings etc mentioned in
the text and also to investigate possible connections with black
hole entropy.

\item{v.}Another point is to study the ${\cal N}=4$ in the large
$N$, fixed $g$ limit on other manifolds or with other gauge
groups, for which we have the exact answer from the topologically
twisted version. This would provide us with some hint about
whether this phase transition is more general or specific to the
$K3$ case.

\noindent Finally , and in a rather different direction, we think
it would be very interesting to study the 't Hooft large $N$ limit
of the partition function of the ${\cal N}=4$ theory on $K3$.
Presumably there must be some holographic string dual. Since the
gauge theory on $K3$ can be twisted to a topological theory, one
would expect the same thing from the dual string theory. This has
been discussed in
$\HullVG,\LozanoJI,\deMedeirosKX,\NeitzkePF,\DijkgraafTE$. If this
dual string theory can be twisted to a "topological" theory, we
might be able to compute its partition function exactly, as it was
possible for the gauge theory. If all of the above were done then
one would have an exact expression for the partition function on
both sides of an AdS/CFT-like duality and one could test it
exactly in $\lambda$ or even in $N$, in analogy with
$\GopakumarKI$ but for a 4-dimensional gauge theory.

\newsec{Acknowledgements}
I would like to thank  M. Ernebjerg, D. Gaiotto, L. Grant, M.
Guica, N. Iizuka, D. Jafferis, S. Lahiri, J. Marsano, J. Minahan,
L. Motl, G. Pastras, S. Raju, A. Simons, A. Strominger, C. Vafa
for very useful discussions. I would especially like to thank
Shiraz Minwalla for numerous invaluable comments, discussions and
insights. The work of K.P. was supported in part by DOE grant
DE-FG01-91ER40654.

\appendix{A}{Modular forms}

{\bf Definitions}

 The Dedekind eta function $\eta(\tau)$ is
defined by:

\eqn\etadef{\eta(\tau)=q^{1\over 24}
\prod_{m=1}^\infty(1-q^m),\qquad q=\exp(2\pi i \tau).}

It is analytic in the upper half-plane. Under $SL(2,Z)$ it
transforms like:

\eqn\etatrans{\eta(\tau+1)=e^{i{\pi\over 12}}\eta(\tau),\qquad
\eta\left(-{1\over \tau}\right) = (-i\tau)^{\half} \eta(\tau).}

We define:

\eqn\gdefa{G(\tau) = \eta(\tau)^{-24}}

It transforms as:

\eqn\gtrans{G(\tau+1)=G(\tau),\qquad G\left(-{1\over \tau}\right)
= (\tau)^{-12} G(\tau).}

{\bf Asymptotic Expansion}

The function $G(\tau)$ can be written in the form:

\eqn\asymptoticg{G(\tau)=\sum_{n=-1}^\infty d_n q^n,\qquad
q=\exp(2\pi i \tau)} where $d_n$ are integers. Each $d_n$ is equal
to the degeneracy of the level $n+1$ oscillator space of one side
of the bosonic string.

We have that $d_{-1}=1$, $d_0=24$ so for large positive values of
$Im(\tau)$ the function G has the expansion:

\eqn\asymptoticgb{G(\tau) ={1\over q} + 24+ {\cal O}(q)= e^{-2\pi
i \tau}+ 24+{\cal O}\left(e^{2\pi i \tau}\right)} Also it is easy
to show that for large $n$ we have a Hagedorn-like growth for
$d_n$ : \eqn\hagedorn{d_n \sim \exp(4\pi \sqrt{n})}

\appendix{B}{Partition function of ${\cal N}=4$ on $K3$ at large $N$}
In this appendix we will study the large $N$ limit of the
partition function $\partitionhecke$, for $\tau$ in the first
fundamental domain ${\cal F}_o$ of $SL(2,Z)$. Our goal is to
identify the largest term in the sum, let us call this term $F$,
as well as the second largest, call it $S$. We then show that the
ratio of $F/S$ grows exponentially in the large $N$ limit. Since
the number of terms in the partition function grows at most like
some power of $N$, we conclude that $F$ contributes exponentially
more than all the other terms together, and thus it is reliable to
study the large $N$, fixed $\tau$ limit of the partition function
by looking only at this largest term $F$.

{\bf Partition function for $\theta=0$} In the next subsection
we present a shorter argument that can prove our result even for
$\theta=0$. In this subsection we develop a longer but more
explicit proof for the case $\theta=0$.

For $\theta=0$, the coupling $\tau=i {4\pi\over g^2}$ is purely
imaginary. We now show that for any $a,b\in\Re$ and $a>0$:

\eqn\bapa{|G(ia)|\geq |G(ia+b)|}

From the definition of G:

\eqn\gdefa{G(\tau) = {1\over q} \prod_{m=1}^\infty
(1-q^m)^{-24},\qquad q=exp(2\pi i \tau)} we have:

\eqn\bapb{\left|{G(ia)\over G(ia+b)}\right| = \prod_{m=1}^\infty
\left|{ 1-e^{2\pi i b m}e^{-2\pi a m}\over 1-e^{-2\pi a
m}}\right|^{24}\geq 1} since every factor in the product is
independently greater than or equal to one.

So in the partition function:

\eqn\partitionheckeap{Z(\tau) ={1\over N^3} \sum_{0\leq a,b,d \in
Z ad=N; b\leq d-1}d\, G\left({a \tau + b\over d}\right).} for the
case $\theta=0$, it is true that for given $a$ and $d$, the term
that will contribute the most is the one with $b=0$. So, we only
need to compare the terms $G\left({\a\over d}\tau\right)$ for
$a,d\in\aleph$ and $a d = N$. The possible values of $a$ and $d$
are the divisors of $N$. Let us assume that $N$ is even and look
at the asymptotic expansion of the various terms for large $N$,
for imaginary $\tau=i \tau_2$:

\eqn\bapc{\matrix{&a=N,d=1:\qquad\qquad & G(N\tau) \sim & exp(N 2
\pi \tau_2)\cr \cr& a=N/2,d=2:\qquad\qquad & G(N/4\tau) \sim &
exp\left({N\over 4} 2 \pi \tau_2\right)\cr \cr &...&...&...\cr\cr&
a=2,d=N/2:\qquad\qquad & G\left({4\over N}\tau\right) \sim
G\left(-{N\over 4}{1\over \tau}\right) \sim & exp\left({N\over 4}
2 \pi {1\over \tau_2}\right) \cr \cr & a=1,d=N:\qquad\qquad &
G\left({1\over N}\tau\right) \sim G\left(-N{1\over \tau}\right)
\sim & exp\left(N 2 \pi {1\over \tau_2}\right)}}

It is clear that it is either the term $G(N\tau)$ or
$G\left({\tau\over N}\right)$ that will be the largest for large
enough $N$. Specifically in the first fundamental domain  ${\cal
F}_o$ and for $\theta=0$ we have $\tau_2>1$ so $G(N\tau)$ is the
largest. We also see that the second largest term (no matter which
one it is) is exponentially suppressed compared to $G(N\tau)$ (for
$\tau_2>1$). Similarly for $\tau_2<1$ the dominant term is
$G\left({\tau\over N}\right)$. It is clear that the same will
happen with any other divisor of $N$. This completes our analysis
for $\theta=0$.

{\bf Partition function for $\theta\neq 0$}

Now we claim that for $\tau$ complex and in the first fundamental
domain ${\cal F}_o$ of $SL(2,Z)$ it is still $G(N\tau)$ that is
the largest term.

As we take $N\ra \infty$ there are some terms in the partition
function, for which the argument of the function $G$ has the
largest imaginary part, and which goes to infinity $N\ra\infty$.
These are the terms for which $a>d$. From them the term $G(N\tau)$
is clearly the dominant one. There are also terms that have the
smallest imaginary part, which goes to zero as $N\ra\infty$. These
are the terms for which $a<d$. For them we do a modular
transformation and then they get a very large and positive
imaginary part that goes like: ${N a \tau_2\over
a^2\tau_2^2+(a\tau_1 +b)^2 }$. It is easy to show that in the
first fundamental domain the largest of these is the one with
$a=1,\, b=0$, which corresponds to the term $G\left({\tau\over
N}\right)$. In ${\cal F}_o$ we can see that $G(N\tau)$ grows
faster with $N$.

So again we conclude that the largest term in ${\cal F}_o$ is
$G(N\tau)$ and the second-largest term is exponentially smaller.

\appendix{C}{Instanton moduli spaces on $K3$}

In this appendix we will try to find the behavior of the Euler
characteristic $c_{k,N}$ of the $k$-instanton moduli space for
$SU(N)$ on $K3$,  for large $N$ and $k$. Let's assume that $N$ is
prime for simplicity.

The partition function can be written in two different forms. One
is the general form where the contribution from the various
instanton sectors are grouped together:

\eqn\topologicalparta{Z_N(\tau)\sim q^{-N}\sum_{k=0}^\infty
c_{k,N}\, q^k,\qquad q=\exp(2\pi i \tau)} and the other is the
function that was computed in $\VafaTF$:
\eqn\partitionprimea{Z_N(\tau) = {1\over N^3} G(N \tau) +{1\over
N^2} \sum_{m=0}^{N-1} G\left({\tau+m\over N}\right),} where
\eqn\defG{G(\tau)=\eta(\tau)^{-24}  }

Now if we want to compute $c_{k,N}$, according to
$\topologicalparta$ we have to isolate the coefficient of
$q^{k-N}$ in $\partitionprimea$.

$\bullet$ First let's study the sum over $m$ in
$\partitionprimea$. We will use the expansion $\asymptoticg$ for
$G(\tau)$ that we mentioned in Appendix A:

\eqn\asumma{\sum_{m=0}^{N-1} G\left({\tau+m\over N}\right)=
\sum_{k=-1}^\infty d_k q^{k\over N} \sum_{m=0}^{N-1}
\exp\left(2\pi i {m k\over N}\right)} The sum over $m$ forces $k$
to be a multiple of $N$ so we have:

\eqn\asummb{\sum_{m=0}^{N-1} G\left({\tau+m\over N}\right)= N
\sum_{l=0}^\infty d_{l N} q^l}

For $c_{k,N}$ we want the coefficient of $q^{k-N}$. It is equal to
$d_{k N-N^2}$ (up to multiplicative factors that we ignore). Now
using $\hagedorn$ we find:

\eqn\asymptoticEuler{c_{k,N} \sim \exp\left(4\pi \sqrt{k
N-N^2}\right)}

$\bullet$ Now let's look at the term $G(N\tau)$. We have:

\eqn\gntexp{G(N\tau) = \sum_{n=-1}^\infty d_n q^{N n }}

This term can contribute to the $k$-instanton sector only if $k$
is a multiple of $N$ but even then its contribution is $d_{{k\over
N} -1}$ which is suppressed compared to $\asymptoticEuler$, so we
can ignore it in the large $N$ limit.

Thus our final result for the Euler characteristic of the
$k$-instanton moduli space for $SU(N)$ on $K3$ is
$\asymptoticEuler$.

\listrefs
\end